\documentclass{article}

\usepackage{arxiv}

\usepackage[utf8]{inputenc} 
\usepackage[T1]{fontenc}    
\usepackage{hyperref}       
\usepackage{url}            
\usepackage{booktabs}       
\usepackage{amsfonts}       
\usepackage{nicefrac}       
\usepackage{microtype}      
\usepackage{lipsum}		
\usepackage[final]{graphicx}
\usepackage[numbers]{natbib}
\usepackage{doi}
\usepackage{multirow}
\usepackage{float}
\usepackage{subcaption}
\usepackage{placeins}
\usepackage{amsmath}
\usepackage{pifont}
\usepackage[table]{xcolor}

\usepackage{array}
\newcolumntype{L}[1]{>{\raggedright\let\newline\\\arraybackslash\hspace{0pt}}m{#1}}
\newcolumntype{C}[1]{>{\centering\let\newline\\\arraybackslash\hspace{0pt}}m{#1}}
\newcolumntype{R}[1]{>{\raggedleft\let\newline\\\arraybackslash\hspace{0pt}}m{#1}}

\newcommand\linesubsec[1]{\vspace{0.8mm}\noindent\textbf{#1 --- }}
\newcommand{\lesstiny}{\fontsize{7pt}{8pt}\selectfont}

\title{NablAFx: A Framework for Differentiable Black-box and Gray-box Modeling of Audio Effects}

\author{
    Marco Comunità\\
    \texttt{m.comunita@qmul.ac.uk}
    \AND Christian J. Steinmetz \And Joshua D. Reiss
    \AND \\
    Centre for Digital Music\\
    Queen Mary University of London, UK\\
}

\date{}

\renewcommand{\shorttitle}{\textit{arXiv} Template}

\hypersetup{
pdftitle={A template for the arxiv style},
pdfsubject={q-bio.NC, q-bio.QM},
pdfauthor={David S.~Hippocampus, Elias D.~Striatum},
pdfkeywords={First keyword, Second keyword, More},
}

\begin{document}
\maketitle

\begin{abstract}
We present NablAFx, an open-source framework developed to support research in differentiable black-box and gray-box modeling of audio effects. 
Built in PyTorch, NablAFx offers a versatile ecosystem to configure, train, evaluate, and compare various architectural approaches. 
It includes classes to manage model architectures, datasets, and training, along with features to compute and log losses, metrics and media, and plotting functions to facilitate detailed analysis. 
It incorporates implementations of established black-box architectures and conditioning methods, as well as differentiable DSP blocks and controllers, enabling the creation of both parametric and non-parametric gray-box signal chains. The code is accessible at \small{\url{https://github.com/mcomunita/nablafx}}.
\end{abstract}


\section{Introduction}
\label{intro}
Audio effects are central for engineers and musicians to shape timbre, dynamics, and spatialisation of sound~\cite{wilmering2020history}.
Therefore, research related to audio effects, especially with the success of deep learning 
and differentiable digital signal processing (DDSP) \citep{engel2020ddsp}, is a very active field \citep{comunita2024afxresearch}. 
This includes applications such as classification and identification 
\citep{
comunita2021guitar},
parameters estimation 
\citep{
colonel2022direct, 
mitcheltree2023modulation}, 
modeling 
\citep{
comunita2023modelling,
simionato2024comparative},
style transfer \citep{
steinmetz2022style, 
steinmetz2024st}, 
automatic mixing 
\citep{
steinmetz2021automatic, 
sai2023adoption}.
Audio effects modeling is one of the most active applications of differentiable approaches, 
with the majority of methods falling into black-box (i.e., neural networks) and gray-box (i.e., DDSP) paradigms. 
While black-box models achieve state-of-the-art accuracy 
\citep{
wright2019real, 
steinmetz2022efficient, 
comunita2023modelling, 
yeh2024hyper} 
there is interest in gray-box ones 
\citep{
colonel2022reverse,
wright2022grey, 
carson2023differentiable, 
miklanek2023neural,
yeh2024ddsp} 
due to interpretability and potential for efficiency.

Comparing modeling paradigms remains challenging due to significant variations in training and evaluation methods. 
In addition, the lack of standardized implementations for models and DDSP blocks further impedes reproducibility and performance assessment.
There are a growing number of audio effect implementations available to researchers, however existing options remain limited in a number of ways (see Table~\ref{tab:frameworks}).

While the \texttt{Spotify Pedalboard}\footnote{\lesstiny{\url{github.com/spotify/pedalboard}}} library offers Python implementations of common audio effects and allows to define signal chains, these are not differentiable. 
DDSP, introduced in \citep{engel2020ddsp}, provides some differentiable blocks\footnote{\lesstiny{\url{github.com/magenta/ddsp}}}, though they are neither common nor easily reusable, since they are focused on specific applications within audio synthesis.
\texttt{dasp}\footnote{\lesstiny{\url{github.com/csteinmetz1/dasp-pytorch}}}\citep{steinmetz2022style} includes differentiable implementations of common processing and mixing blocks, and while useful when imported into larger projects, the library in not meant to define signal chains.
Interconnections of processors can be defined in \texttt{diffmoog}\footnote{\lesstiny{\url{github.com/aisynth/diffmoog}}}\citep{uzrad2024diffmoog}, although mainly focused on FM synthesis and not suitable for effects modeling.
Also \texttt{GRAFX}\footnote{\lesstiny{\url{github.com/sh-lee97/grafx}}}~\citep{lee2024grafx} enables complex interconnections, but lacks external control, limiting parametric, time-varying, and modulated signal chains for effects modeling.

\texttt{pyneuralfx}\footnote{\lesstiny{\url{github.com/ytsrt66589/pyneuralfx}}}\citep{yeh2024pyneuralfx} is the only framework designed for modeling and, while it includes state-of-the-art neural networks, it focuses only on black-box approaches and does not include time-varying models \citep{comunita2023modelling}. 
Even though it provides functions for inference-time analysis, it lacks logging and plotting features during training and testing. 
Also, experiment configurations are hard to modularize and adapt to different datasets, models, or training procedures, limiting repeatability and comparison.

To address these limitations and advance differentiable audio effects modeling, we propose \textbf{NablAFx}, which provides:
\begin{itemize} 
    \item Black-box architectures, gray-box processors, and controllers for parametric/non-parametric models.
    \item Modules to manage datasets, training, and loss functions.
    \item Tools to log metrics and media during training and testing.
    \item Plotting functions for analysis throughout training.
\end{itemize}

\renewcommand{\arraystretch}{1.1}
\begin{table*}[t]
    \small
    \caption{Python libraries for processing/modeling applications. We show if: they include differentiable (Diff.) implementations, neural networks (NN), DSP processors (Proc.) and controllers (Contr.), they allow to define signal chains and include analysis tools.}
    \label{tab:frameworks}
    \centering
    \begin{tabular}{lcccccc} 
        \hline
        \hline
        Library
        & Diff.
        & NN
        & Proc.
        & Contr.
        & Chains
        & Analysis\\
        \hline
        \texttt{Pedalboard} & \ding{55} & \ding{55} & \ding{51} & \ding{55} & \ding{51} & \ding{55}\\
        \texttt{DDSP} & \ding{51} & \ding{55} & \ding{51} & \ding{55} & \ding{55} & \ding{55} \\
        \texttt{dasp} & \ding{51} & \ding{55} & \ding{51} & \ding{55} & \ding{55} & \ding{55}\\
        \texttt{diffmoog} & \ding{51} & \ding{55} & \ding{51} & \ding{55} & \ding{51} & \ding{55}\\
        \texttt{GRAFX} & \ding{51} & \ding{55} & \ding{51} & \ding{55} & \ding{51} & \ding{55}\\
        \texttt{pyneuralfx} & \ding{51} & \ding{51} & \ding{55} & \ding{55} & \ding{55} & \ding{51}\\
        \hline
        \texttt{NablAFx} & \ding{51} & \ding{51} & \ding{51} & \ding{51} & \ding{51} & \ding{51}\\
        \hline
        \hline
    \end{tabular}
\end{table*}

\section{Framework}
\label{sec:framework}

\begin{figure*}[t]
    \centering
    \includegraphics[width=1\linewidth]{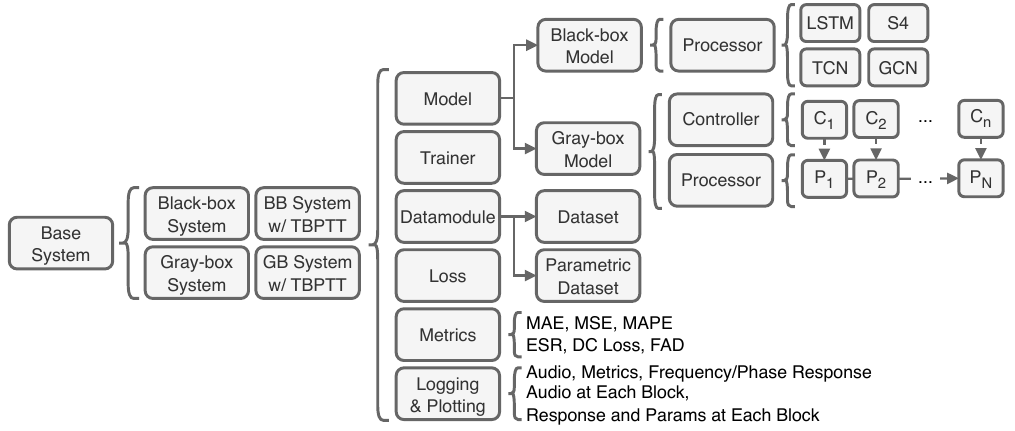}
    \caption{Overview of the NablAFx framework for audio effects modeling}
    \label{fig:framework}
\end{figure*}

NablAFx is a framework for audio effects modeling that allows researchers to easily define, train, evaluate and compare differentiable black-box and gray-box models. 
As shown in Fig.~\ref{fig:framework}, it integrates models, datasets, trainers, loss functions, metrics, and logging/plotting tools. 
Built with PyTorch Lightning\footnote{\lesstiny{\url{lightning.ai/pytorch-lightning}}}, it leverages Weights\&Biases\footnote{\lesstiny{\url{wandb.ai/site}}} to log results and media.

\begin{figure*}[t]
\centering
\begin{subfigure}{.32\textwidth}
    \centering
    \includegraphics[height=3.7cm]{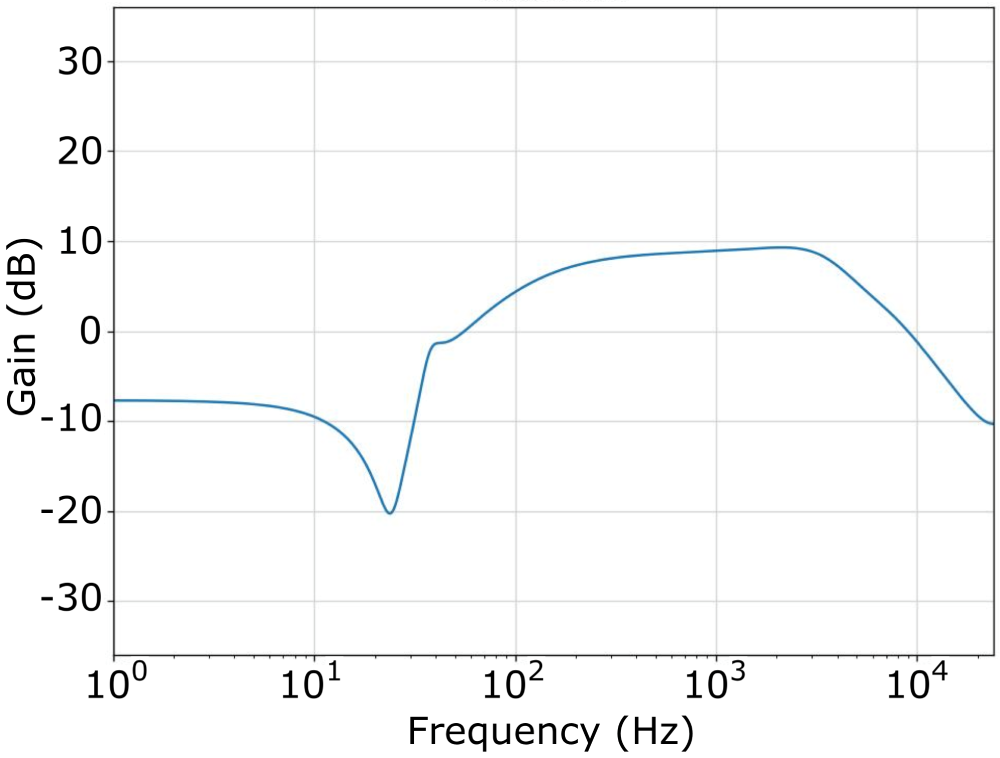}
    \caption{Parametric EQ: frequency response}
    \label{fig:ex_eqblock_response}
\end{subfigure}
\begin{subfigure}{.32\textwidth}
    \centering
    \includegraphics[height=3.7cm]{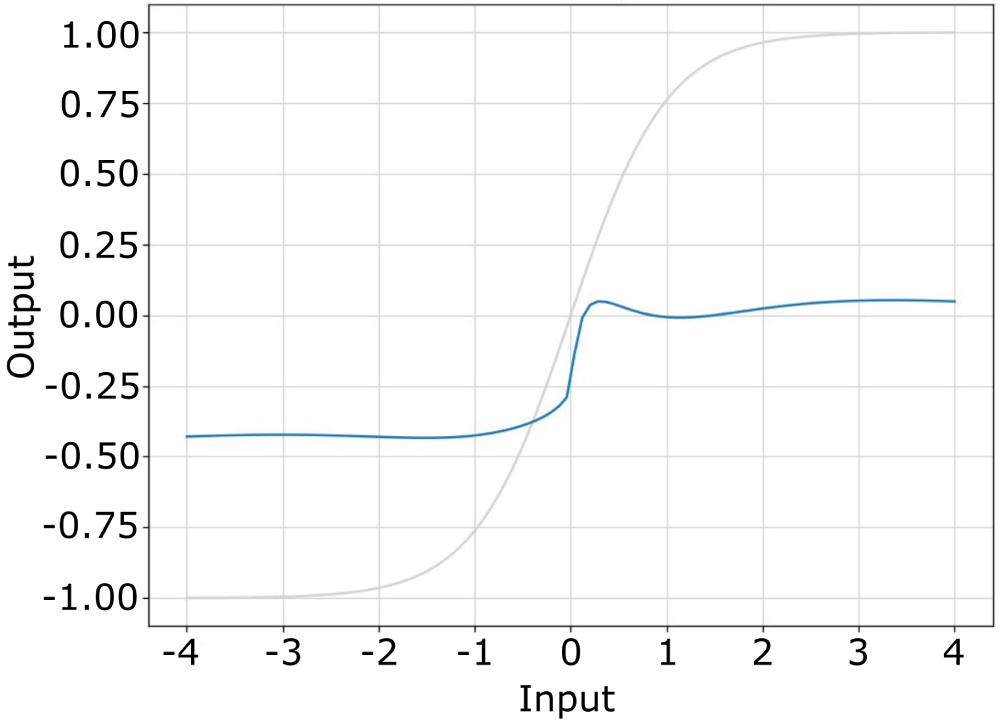}
    \caption{Nonlinearity: amplitude response}
    \label{fig:ex_nonlinblock_response}
\end{subfigure}
\begin{subfigure}{.32\textwidth}
    \centering
    \includegraphics[height=3.7cm]{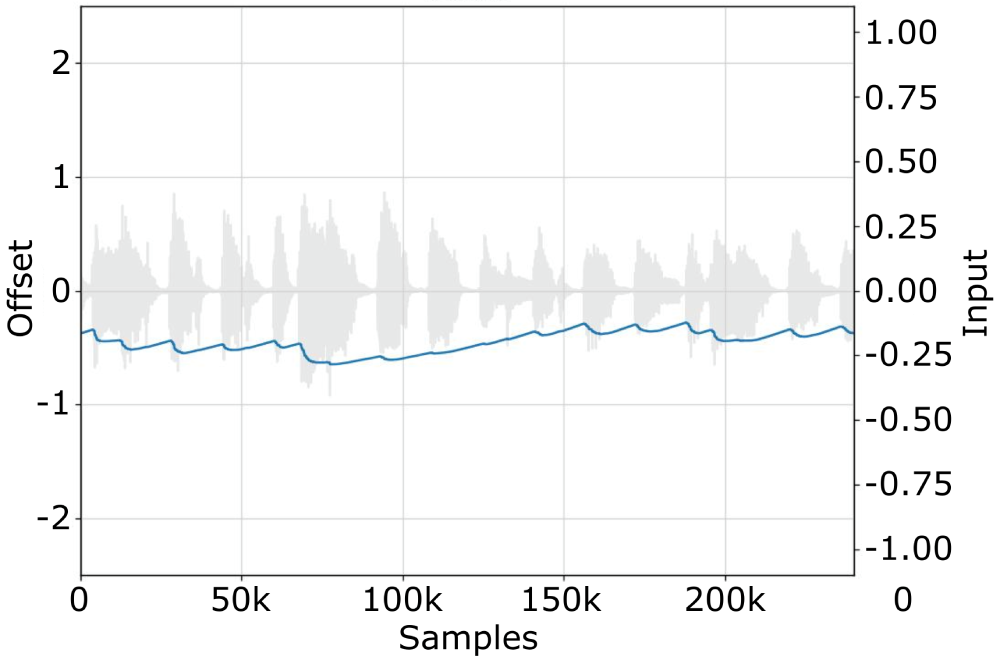}
    \caption{DC offset: time response}
    \label{fig:ex_dcblock_response}
\end{subfigure}
\caption{Examples of plotting features included in NablAFx}
\end{figure*}

\linesubsec{System}
In NablAFx all necessary functionalities are contained in an audio effects modeling system class. 
The \textit{BaseSystem} class handles the initialization of loss functions, optimizers, learning rate scheduler, metrics, and includes shared methods to compute and log loss, metrics, audio and frequency/phase response.
The \textit{BaseSystem} is divided into \textit{BlackBoxSystem} and \textit{GrayBoxSystem}, which initialize black-box and gray-box models, respectively, and implement train, validation, and test steps. 
The \textit{GrayBoxSystem} adds methods to log audio output, plot/log frequency and time responses, and parameters values for each stage of the signal chain. 
Both systems are extended with \textit{WithTBPTT} classes, which implement truncated backpropagation through time to enable faster training of recurrent networks\citep{wright2019real}.

\linesubsec{Model}
In our framework, black-box models can be any neural network - with outputs defined as a function of input and controls $y = f(x,c)$ -  represented by the \textit{Processor} class in \textit{BlackBoxModel}. 
Gray-box models comprise interconnected differentiable blocks, forming a function composition: $y = (f_{1} \circ f_{2} \circ \ldots \circ f_{N})(x,c)$, and the \textit{Processor} class defines a chain of processors.
A \textit{Controller} class defines a chain of controllers, each associated with a processor, allowing the definition of parametric and time-varying models that are a function of both input audio and controls.

\linesubsec{Data}
The \textit{DataModule} class takes care of initializing the dataset and dataloaders for train, validation and testing. \textit{AudioEffectDataset} and \textit{ParametricAudioEffectDataset} classes are used to manage data for non-parametric and parametric models. 

\linesubsec{Metrics}
Metrics are computed with: 
\textit{torchmetrics}\footnote{\lesstiny{\url{lightning.ai/docs/torchmetrics/stable/}}} for mean absolute error (MAE), mean squared error (MSE) and mean absolute percentage error (MAPE); \textit{auraloss}\footnote{\lesstiny{\url{github.com/csteinmetz1/auraloss}}}\citep{steinmetz2020auraloss} for error-to-signal ratio (ESR) 
and DC loss; and the 
\textit{frechet-audio-distance}\footnote{\lesstiny{\url{github.com/gudgud96/frechet-audio-distance}}} package for Frechét Audio Distance (FAD) \citep{kilgour2018fr}.

\linesubsec{Plotting}
In addition to logging losses, metrics and audio examples, we provide methods to plot and log frequency/phase response for the whole system, as well as frequency/time response and parameters values for each DDSP block in a gray-box system.
We offer two methods to compute the frequency and phase response: one using an exponential sine sweep\footnote{\lesstiny{\url{ant-novak.com/pages/sss/}}} \citep{novak2009nonlinear}, suitable for linear and mildly nonlinear systems, and a custom method designed for nonlinear systems.
The latter measures the system's response in steps, using sinusoidal inputs at exponentially spaced frequencies.
To ensure reliable measurements, each sinusoid lasts several seconds for the system to reach steady state, with magnitude/phase response computed only from the final segment.
\begin{align*}
    x = x[-T* \lfloor f_{s}/f_{1} \rfloor:] \\ 
    y = y[-T* \lfloor f_{s}/f_{1} \rfloor:]
\end{align*}
where $T$ is the signal duration (e.g. 5~s), $f_{s}$ is the sample rate and $f_{1}$ is the minimum frequency of the stepped sweep (e.g., 10~Hz). 
Fig.~\ref{fig:ex_eqblock_response} shows the frequency response of a Parametric EQ block, while Fig.~\ref{fig:ex_nonlinblock_response} and \ref{fig:ex_dcblock_response} display examples of learned nonlinearity (vs. $tanh$, light gray) and time-varying DC offset (vs. input signal, light gray).

\begin{figure*}[t]
\centering
\begin{subfigure}{.125\textwidth}
  \centering
  \includegraphics[width=2cm]{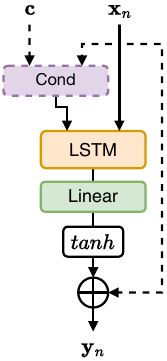}
  \caption{LSTM}
  \label{fig:lstm-block-diag}
\end{subfigure}
\begin{subfigure}{.195\textwidth}
  \centering
  \includegraphics[width=2.5cm]{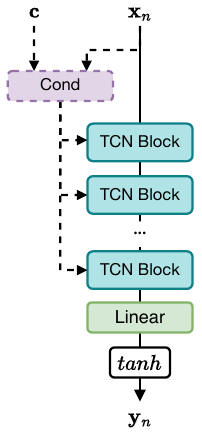}
  \caption{TCN}
  \label{fig:tcn-block-diag}
\end{subfigure}
\begin{subfigure}{.195\textwidth}
  \centering
  \includegraphics[width=3cm]{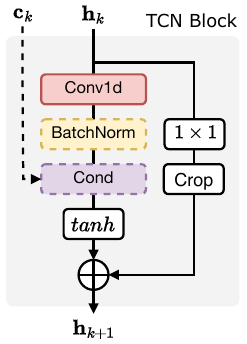}
  \caption{TCN block}
  \label{fig:tcnblock-block-diag}
\end{subfigure}
\begin{subfigure}{.265\textwidth}
  \centering
  \includegraphics[width=3.8cm]{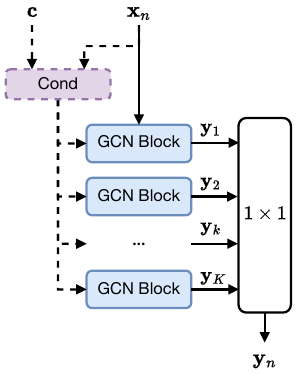}
  \caption{GCN}
  \label{fig:gcn-block-diag}
\end{subfigure}
\begin{subfigure}{.195\textwidth}
  \centering
  \includegraphics[width=3.05cm]{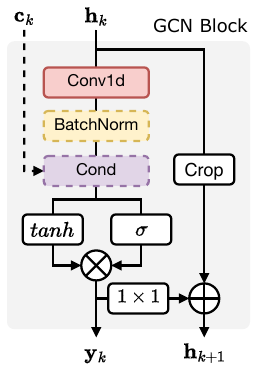}
  \caption{GCN block}
  \label{fig:gcnblock-block-diag}
\end{subfigure}
\caption{Black-box architectures included in NablAFx}
\label{fig:architectures1}
\end{figure*}

\subsection{Differentiable Black-box Models}
\label{sec:bb-models}
This section provides an overview of state-of-the-art neural network architectures and conditioning methods included in NablAFx.

\linesubsec{LSTM}
The recurrent neural network architecture we implement is widely adopted for nonlinear effects (e.g., overdrive, distortion, guitar amps) \citep{wright2019real, wright2020real}, nonlinear time-varying effects (e.g., fuzz, compressor) \citep{steinmetz2022efficient, comunita2023modelling}, and modulation effects (e.g., phaser, flanger) \citep{wright2021neural, mitcheltree2023modulation}. 
As shown in Fig.~\ref{fig:lstm-block-diag}, it consists of a single LSTM layer, a linear layer, and a $tanh$ activation. 
For parametric models, a conditioning block processes control values and optionally the input sequence.

\linesubsec{TCN}
Temporal Convolution Networks (TCNs), introduced in \citep{lea2016temporal} and shown to outperform recurrent architectures \citep{bai2018empirical} on a variety of tasks, were proposed for audio effects modeling \citep{steinmetz2020learning, stein2010automatic, steinmetz2022efficient} and applied to linear (EQ, reverb) and nonlinear time-varying (compressor) effects.
The architecture (Fig.~\ref{fig:tcn-block-diag}) consists of a series of residual blocks (Fig.~\ref{fig:tcnblock-block-diag}) made of 1-dimensional convolutions with increasing dilation factors, optionally followed by batch normalization and conditioning block, and an activation function (here $tanh$). 
A linear layer matches the output channels to the input size.

\linesubsec{GCN}
Gated Convolution Networks (GCNs), introduced in \citep{rethage2018wavenet} as a feed-forward WaveNet, 
are a special case of TCNs with gated convolutions. 
GCNs have been used in 
\citep{
damskagg2019real, 
wright2020real} 
for nonlinear audio effects (guitar amp, overdrive, distortion) and in \citep{comunita2023modelling} for nonlinear time-varying effects (compressor, fuzz).
Beside the activation function at each block (Fig.\ref{fig:gcnblock-block-diag}), a GCN (Fig.\ref{fig:gcn-block-diag}) differs from a TCN in that its output is a linear combination of the activation features at each block.

\linesubsec{S4}
Structured state space sequence models (S4) were introduced in \citep{gu2021efficiently} as a general sequence modeling architecture and shown to outperform recurrent, convolutional and Transformer architectures on a variety of tasks. 
An S4 layer is a differentiable implementation of an infinite impulse response (IIR) system in state-space form, with a theoretically infinite receptive field, similar to recurrent networks.
Based on these observations state-space models were adopted for non-linear time-varying (compressor) effects modeling~\citep{yin2024modeling, simionato2024modeling}.

\begin{figure*}[t]
\centering
\begin{subfigure}{.245\textwidth}
  \centering
  \includegraphics[width=2.6cm]{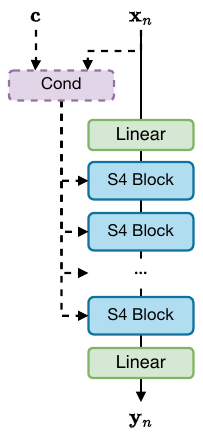}
  \caption{S4}
  \label{fig:s4-block-diag}
\end{subfigure}
\begin{subfigure}{.245\textwidth}
  \centering
  \includegraphics[width=3.2cm]{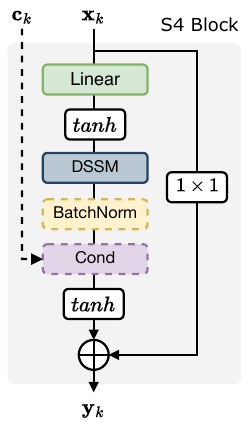}
  \caption{S4 block}
  \label{fig:s4block-block-diag}
\end{subfigure}
\begin{subfigure}{.245\textwidth}
  \centering
  \includegraphics[width=3cm]{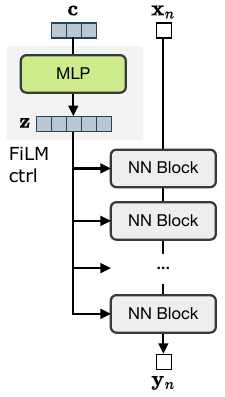}
  \caption{FiLM controller}
  \label{fig:film_ctrl}
\end{subfigure}
\begin{subfigure}{.245\textwidth}
  \centering
  \includegraphics[width=4cm]{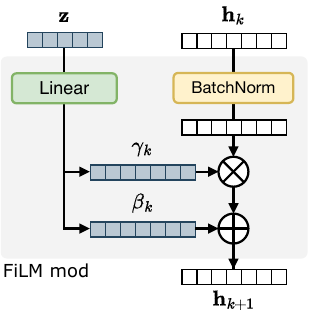}
  \caption{FiLM modulator}
  \label{fig:film_mod}
\end{subfigure}
\caption{Black-box architectures and conditioning methods included in NablAFx}
\label{fig:architectures2}
\end{figure*}

The architecture in our framework, based on \citep{yin2024modeling} (Fig.~\ref{fig:s4-block-diag}), consists of S4 blocks.
Unlike standard convolutional ones, S4 layers are not combined or mixed across data channels, this explains the use of a linear layer and activation function ($\tanh$) at the input of each S4 block (Fig.~\ref{fig:s4block-block-diag}) for affine transformations along the channel dimension. 
These are followed by an S4D layer \citep{gupta2022diagonal}, which uses diagonal matrices for a parameter-efficient implementation, optional batch normalization and conditioning block, followed by an activation function ($tanh$ in this case).
Linear layers are used at the start and end to adjust the channel count to match the input data.

\subsubsection{Conditioning for Black-box Models}
\label{sec:cond-bb-models}
Conditioning mechanisms for black-box models have been explored for different purposes: to include parametric control \citep{
steinmetz2022efficient, 
simionato2024comparative, 
yin2024modeling}, 
to capture long-range dependencies \citep{comunita2023modelling} or for modulation in LFO-driven effects \citep{mitcheltree2023modulation, wright2021neural}.
While concatenation and feature-wise linear modulation (FiLM) \citep{
steinmetz2022efficient, 
yin2024modeling, 
mitcheltree2023modulation, 
simionato2024comparative} 
remain the most common methods, 
temporal FiLM (TFiLM) has been adopted to capture time-varying behavior \citep{comunita2023modelling}.
Beside these established methods, in this work we also propose three further conditioning methods: time-varying concatenation (TVCond), tiny TFiLM (TTFiLM) and time-varying FiLM (TVFiLM), as efficient implementations of time-varying conditioning similar to TFiLM. 
Concatenating control values ($\mathbf{c}$) to the input sequence ($\mathbf{x_{n}}$) along the channel dimension is a simple, parameter-efficient conditioning method. 
It serves as a baseline for recurrent networks and has been used for parametric control in, e.g., compression~\citep{steinmetz2022efficient} and overdrive~\citep{yeh2024hyper}.

Equally common is FiLM conditioning, mainly when using TCN \citep{steinmetz2022efficient, yeh2024hyper}, GCN \citep{yeh2024hyper} or S4 \citep{yin2024modeling, simionato2024modeling} backbones, with works adopting it for compressors \citep{steinmetz2022efficient, simionato2024modeling, simionato2024comparative} and overdrive \citep{yeh2024hyper, simionato2024comparative} modeling.
Introduced in \citep{perez2018film} as a general-purpose conditioning method, FiLM modulates a neural network's intermediate features using a conditioning vector $\mathbf{c}$. It learns functions $f$ and $g$ to generate scaling ($\gamma_{k,c}=f(\mathbf{c})$) and bias ($\beta_{k,c}=g(\mathbf{c})$) parameters for each layer $k$ and channel $c$, which are used to modulate the activations at each layer $\mathbf{h}_{k,c}$, via a feature-wise affine transformation:
\begin{equation}
    \text{FiLM}(\mathbf{h}_{k,c},\gamma_{k,c},\beta_{k,c})=\gamma_{k,c}\cdot\mathbf{h}_{k,c}+\beta_{k,c}.
\end{equation}
In practice, $f$ and $g$ are neural networks (Fig.~\ref{fig:film_ctrl}) that learn a latent representation $\mathbf{z}$ of the conditioning vector $\mathbf{c}$; then, a linear layer uses the latent representation to generate scaling and bias parameters for each block of the main network (Fig.~\ref{fig:film_mod}).

\begin{figure*}[t]
\centering
\begin{subfigure}{.26\textwidth}
  \centering
  \includegraphics[height=5.5cm]{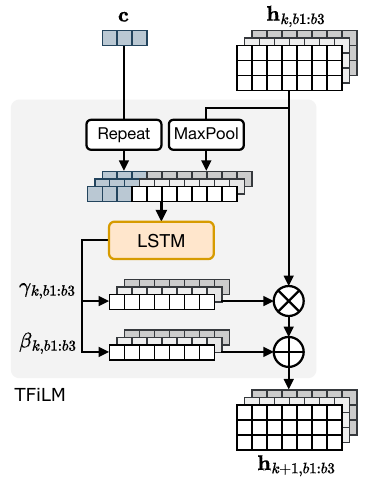}
  \caption{TFiLM}
  \label{fig:tfilm}
\end{subfigure}
\begin{subfigure}{.245\textwidth}
  \centering
  \includegraphics[width=4.2cm]{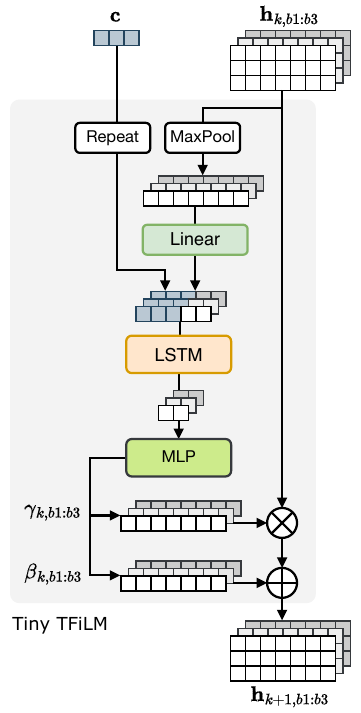}
  \caption{TTFiLM}
  \label{fig:ttfilm}
\end{subfigure}
\begin{subfigure}{.235\textwidth}
  \centering
  \includegraphics[width=3.5cm]{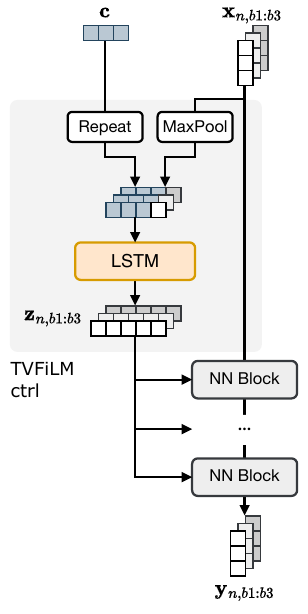}
  \caption{TVFiLM controller}
  \label{fig:tvfilm_ctrl}
\end{subfigure}
\begin{subfigure}{.24\textwidth}
  \centering
  \includegraphics[height=4.5cm]{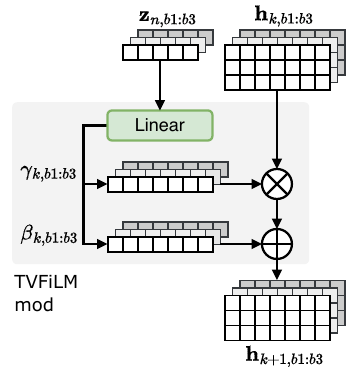}
  \caption{TVFiLM modulator}
  \label{fig:tvfilm_mod}
\end{subfigure}
\caption{Black-box architectures and conditioning methods included in NablAFx}
\label{fig:test}
\end{figure*}

TFiLM \citep{birnbaum2019temporal} enhances network expressivity by using recurrent networks to modulate intermediate features over time as a function of layer activations $\mathbf{h}_{k}$ and optionally a conditioning vector $\mathbf{c}$ (Fig.~\ref{fig:tfilm}).
Given a sequence of activations $\mathbf{h}_k$ from the $k$-th block of a network, the sequence is split into $T$ blocks of $B$ samples $\mathbf{h}_{k,b_{1}:b_{T}}$ along the sequence dimension. 
For each block $\mathbf{h}_{k,b_{t}}$, 1-dimensional max pooling downsamples the signal by a factor of $B$. 
To include the conditioning vector $\mathbf{c}$, it is repeated $T$ times and concatenated with the downsampled activations.
Then, an LSTM generates scaling $\gamma_{k,b_{1}:b_{T},c}$ and bias $\beta_{k,b_{1}:b_{T},c}$ parameters for each channel $c$, which are used to modulate the activations in each block via an affine transformation:
\begin{equation*}
    \begin{split}
        \text{TFiLM}(\mathbf{h}_{k,b_{1}:b_{T},c},\gamma_{k,b_{1}:b_{T},c},\beta_{k,b_{1}:b_{T},c})=\\
        \gamma_{k,b_{1}:b_{T},c}\cdot\mathbf{h}_{k,b_{1}:b_{T},c}+\beta_{k,b_{1}:b_{T},c}.
    \end{split}
\end{equation*}
In its standard formulation, TFiLM conditioning adds a recurrent network for each block in the main neural network, which can lead to a significant increase in parameters due to the number of blocks (typically 5-10) and channels (typically 16-32). 

To retain TFiLM's expressivity while reducing parameters and computational cost, we propose two methods: TTFiLM and TVFiLM.
TTFiLM (Fig.\ref{fig:ttfilm}) is structurally similar to TFiLM, and reduces the computational complexity by using fewer channels in the recurrent network, achieved through a linear layer before it. The output is then scaled up to the required number of scaling $\gamma_{k,b_{1}:b_{T},c}$ and bias $\beta_{k,b_{1}:b_{T},c}$ channels using a small MLP.
TVFiLM is a time-varying extension of FiLM conditioning. 
It replaces the MLP in the FiLM controller (Fig.~\ref{fig:film_ctrl}) with a recurrent network (Fig.~\ref{fig:tvfilm_ctrl}), creating a time-dependent latent representation $\mathbf{z}_{n,b{1}:b_{T}}$ shared across the main network's blocks. 
Modulation sequences are then generated at each block via a linear layer (Fig.\ref{fig:tvfilm_mod}), similarly to standard FiLM (Fig.\ref{fig:film_mod}).

We also implement time-varying concatenation (TVCond) for recurrent models by using a TVFiLM controller to generate a time-dependent conditioning sequence, which is concatenated to the input for greater expressivity compared to standard concatenation.

\subsection{Differentiable Gray-box Models}
\label{sec:gb-models}
As described in Sec.~\ref{sec:framework} we define a gray-box model as a sequence of differentiable processors, each with an associated controller which generates the control parameters that dictate the behavior of the processor. 

\subsubsection{Differentiable Audio Processors}
\label{sec:ddsp-processors}
For our application, we define three types of audio processors: basic (e.g., phase inversion, gain), filters (e.g., EQ, shelving), and nonlinearities (e.g., tanh, MLP). 
Most processors can be controlled by any of the controllers in Sec.~\ref{sec:ddsp-controllers}, enabling parametric and time-varying configurations.
All filters are implemented as differentiable biquads \citep{nercessian2020neural},
which are defined by a second order difference equation:
\begin{equation*}
    \begin{split}
        y[n]=\frac{1}{a_{0}}(b_{0}x[n]+b_{1}x[n-1]+b_{2}x[n-2]+\\
        -a_{1}y[n-1]-a_{2}y[n-1])    
    \end{split}
\end{equation*}
with the transfer function:
\begin{equation*}
    H(z)=\frac{b_{0}+b_{1}z^{-1}+b_{2}z^{-2}}{a_{0}+a_{1}z^{-1}+a_{2}z^{-2}}.
\end{equation*}
Biquad coefficients are calculated based on center/cutoff frequency (Hz), gain (dB), and Q factor, following Robert Bristow-Johnson's method\footnote{\lesstiny{\url{www.musicdsp.org/en/latest/Filters/197-rbj-audio-eq-cookbook.html}}}.
To implement \textit{N}\textsuperscript{th} order filters (e.g., Parametric EQ) we follow the common practice of using \textit{K} cascaded second order sections:
\begin{equation}\label{eq:hz}
    H(z)=\prod_{k=0}^{K}H_{k}(z)
\end{equation}
The frequency response is computed evaluating the transfer function along the unit circle in the complex plane and taking the magnitude:
\begin{equation*}
    \left|H(e^{j\omega})\right|=\left|\prod_{k=0}^{K}H_{k}(e^{j\omega})\right|.
\end{equation*}
For efficiency, during training we adopt the frequency sampling method, which approximates a cascade of second order IIR filters by computing the frequency response as in Eq.~\ref{eq:hz}, applying the convolution in the frequency domain via multiplication and using the inverse FFT to transform the signal back to the time domain:
\begin{equation*}
    y[n] = F^{-1}[Y(z)] = F^{-1}[X(z)H(z)].
\end{equation*}
In the following paragraph we describe each processor. Generally, all processors' parameters can depend on the input $\mathbf{x}$ and/or controls $\mathbf{c}$. 
For simplicity, we omit time dependency in the notation (e.g., $\textnormal{Gain}_\textnormal{{dB}}$ instead of $\textnormal{Gain}_\textnormal{{dB}}[n]$), but with the controllers in Section~\ref{sec:ddsp-controllers}, these can be made to change over time.

\linesubsec{Phase Inversion} Invert the phase of the input. 
\begin{equation*}
    y[n]=-x[n].
\end{equation*}

\linesubsec{Gain} Multiply input by a gain value in dB.
\begin{equation*}
    y[n]=x[n]*10^{(\textnormal{Gain}_\textnormal{{dB}}/20)}.
\end{equation*}

\linesubsec{DC Offset} Add a constant value to the input.
\begin{equation*}
    y[n]=x[n]+O
\end{equation*}

\linesubsec{Lowpass/Highpass} Second order lowpass/highpass implemented as a single biquad section.
\begin{alignat*}{2}
    b_{0} &= (1 \mp \cos\omega_{0}) / 2  & \quad & a_{0} = 1 + \alpha \\
    b_{1} &= \pm (1 \mp \cos\omega_{0})  & \quad & a_{1} = -2 * \cos\omega_{0} \\
    b_{2} &= (1 \mp \cos\omega_{0}) / 2  & \quad & a_{2} = 1 - \alpha
\end{alignat*}
where $\omega_{0} = 2\pi(f_{0}/f_{s})$ with $f_{0}$ and $f_{s}$ cutoff/sampling frequency, and $\alpha = \sin\omega_{0}/(2Q)$. 
Each filter is defined by 2 parameters: cutoff frequency and Q factor.

\linesubsec{Low/High Shelf} Second order low/high shelving filter implemented as a single biquad section.
\begin{equation*}
    \begin{split}
        b_{0} &= A [(A + 1) \mp (A - 1) \cos\omega_{0} + 2 \sqrt{A} \alpha] \\    
        b_{1} &= \pm2A[(A - 1) \mp (A + 1) \cos\omega_{0}] \\
        b_{2} &= A[(A + 1) \mp (A - 1) \cos\omega_{0} \mp 2\sqrt{A}\alpha] \\
        a_{0} &= (A + 1) \pm (A - 1)\cos\omega_{0} + 2\sqrt{A}\alpha \\
        a_{1} &= \mp2[(A - 1) \pm (A + 1) \cos\omega_{0}] \\
        a_{2} &= (A + 1) \pm (A - 1) \cos\omega_{0} - 2\sqrt{A}\alpha
    \end{split}
\end{equation*}
where $A=10^{(Gain_{dB}/40)}$.
Each filter is defined by 3 parameters: gain, cutoff frequency, Q factor.

\linesubsec{Peak/Notch} Second order peak/notch filter implemented as a single biquad section.
\begin{alignat*}{2}
    b_{0} &= 1 + \alpha A       & \quad & a_{0} = 1 + \alpha / A \\
    b_{1} &= -2 \cos\omega_{0}  & \quad & a_{1} = -2\cos\omega_{0} \\
    b_{2} &= 1 - \alpha A       & \quad & a_{2} = 1 - \alpha / A
\end{alignat*}
Each filter is defined by 3 parameters: gain, center frequency, Q factor.

\linesubsec{Parametric EQ} We define a Parametric EQ as a chain of 5 filters: low shelf, three peak/notch filters and a high shelf. 
Each Parametric EQ has 15 parameters.

\linesubsec{Shelving EQ} We define a Shelving EQ as a chain of 4 filters: highpass, low shelf, high shelf, lowpass. 
Each Shelving EQ is defined by a total of 10 parameters.

\linesubsec{Static FIR Filter} We define a Static FIR Filter using a SIREN layer~\citep{sitzmann2020implicit}, which stores the tap values of an $N^{th}$-order impulse response.
\begin{equation*}
    y[n]=b_{0}x[n]+b_{1}x[n-1]+\ldots+b_{N}x[n-N]
\end{equation*}
The network can be initialized with a pre-trained response (e.g., loudspeaker) and its hyperparameters (hidden dimension and layers count) to be customized.

\linesubsec{Tanh Nonlinearity} Standard hyperbolic tangent.

\linesubsec{Static MLP Nonlinearity} MLP Nonlinearity implemented with SIREN layer, initialized by default with a pre-trained model approximating a $tanh$.

\linesubsec{Static Rational Nonlinearity} 
A Padé approximant \citep{baker1961pade} is a rational function of order $[m/n]$ that best approximates a function $f(x)$ near a specific point, with $m \geq 0$ and $n \geq 1$:
\begin{equation*}
    R(x)=\frac{a_{0}+a_{1}x+a_{2}x^{2}+\cdots+a_{m}x^{m}}{1+b_{1}x+b_{2}x^{2}+\cdots+b_{n}x^{n}}
\end{equation*}
which agrees with $f(x)$ to the highest possible order. 
which amounts to:
\begin{alignat*}{1}
    f(0) &= R(0) \\
    f'(0) &= R'(0) \\
    \vdots \\
    f^{(m+n)} &= R^{(m+n)}(0).
\end{alignat*}
Learnable Padé approximants\footnote{\lesstiny{\url{github.com/ml-research/rational_activations}}} \citep{molina2019pad} enable flexible rational activation functions with few weights (numerator and denominator coefficients). 
We define a learnable Static Rational Nonlinearity using a single rational activation layer, initialized by default to a $tanh$ approximation of order $[6,5]$.

\subsubsection{Differentiable Controllers}
\label{sec:ddsp-controllers}
\begin{figure}[t]
    \centering
    \includegraphics[width=.6\linewidth]{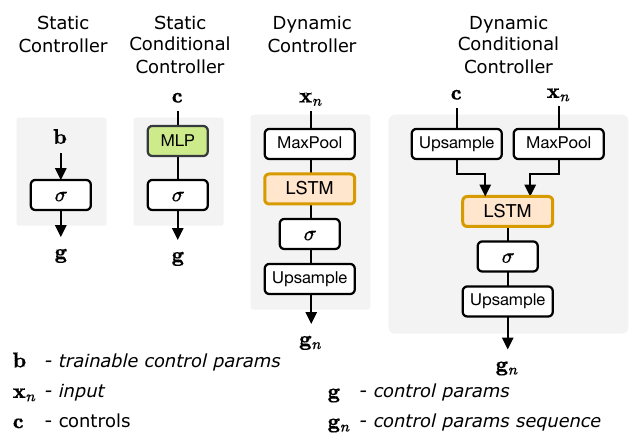}
    \caption{Controllers included in NablAFx}
    \label{fig:controllers}
\end{figure}

We define five types of differentiable controllers (Fig.~\ref{fig:controllers}) to generate control parameters for an audio processor. 

\linesubsec{Dummy} 
A Dummy controller is a placeholder for processors that don't require control parameters (e.g., Phase Inversion, Static FIR Filter, Static Nonlinearity).

\linesubsec{Static} A Static controller is a tensor of trainable controls $\mathbf{b}$ - one for each control parameter in the respective processor - followed by a sigmoid function to limit the output to the [0,1] range: $g = \sigma(\mathbf{b})$.

\linesubsec{Static Conditional} 
A Static Conditional controller uses an MLP with a sigmoid activation to adjust control parameters based on audio effects controls (or some other fixed values): $g = f(\mathbf{c})$. Hyperparameters include number of input controls and output control parameters, number of layers, and hidden dimensions.

\linesubsec{Dynamic} 
A Dynamic controller is used to adjust control parameters over time based on another signal, oftentimes the input audio: $g[n]=f(x[n])$. 
The control signal is downsampled (default downsampling factor is 128), processed through an LSTM, a sigmoid activation, and upsampled to output a control parameters sequence $\mathbf{g}_{n}$ at the original rate. Hyperparameters include block size (i.e., downsampling factor) and number of recurrent layers, while the hidden size is set by the number of control parameters for each processor.

\linesubsec{Dynamic Conditional} 
A Dynamic Conditional controller adjusts control parameters based on both fixed values (typ., audio effect controls) and a time-varying control (typ., input signal): $g[n]=f(x[n],c)$. The signal is downsampled while the controls are upsampled and concatenated, the sequence processed by an LSTM, and after sigmoid activation and upsampling, the control parameters sequence $\mathbf{g}_{n}$ is returned at the original rate.

Although the control parameters sequences are at sampling rate, the block size (i.e., downsampling rate) is used internally in each processors to downsample the sequence so that the coefficients are recomputed once per block. 
This is not a limitation, as setting the block size to 1 provides sequences at audio rate. 
No interpolation methods have been implemented for smooth control sequences at the time of writing.

\setlength{\tabcolsep}{0pt}
\renewcommand{\arraystretch}{0.95}
\setcounter{table}{1}
\begin{table*}[b]
    \small
    \caption{Parametric models included in the experiments. Cond. = conditioning method, R.F. = receptive field in samples.
    PEQ = Parametric EQ, G = Gain, O = Offset, MLP = Multilayer Perceptron, RNL = Rational Non Linearity. Controllers: 
    .s = static, .d = dynamic, .sc = static conditional, .dc = dynamic conditional}
    \label{tab:models}
    \centerline{
        \begin{tabular}{L{2.8cm}C{1.3cm}R{1.1cm}C{1.1cm}C{1.1cm}C{1.3cm}C{1.5cm}R{1.4cm}R{1.3cm}R{1.3cm}}
            \hline
            \hline
            Model
                & Cond.
                    & R.F.
                        & Blocks
                            & Kernel
                                & Dilation
                                    & Channels
                                        & \# Params 
                                            & FLOP/s 
                                                & MAC/s\\ 
            \hline
            TCN-F-45-S-16 & FiLM & 2047 & 5 & 7 & 4 & 16 & 15.0k & 736.5M & 364.3M\\
            TCN-TF-45-S-16 & TFiLM & 2047 & 5 & 7 & 4 & 16 & 42.0k & 762.8M & 364.2M\\
            TCN-TTF-45-S-16 & TTFiLM & 2047 & 5 & 7 & 4 & 16 & 17.3k & 744.0M & 367.4M\\
            TCN-TVF-45-S-16 & TVFiLM & 2047 & 5 & 7 & 4 & 16 & 17.7k & 740.4M & 366.2M\\
            \hline
            \hline
        \end{tabular}
    }
    \centerline{
        \begin{tabular}{L{2.8cm}C{1.3cm}R{1.1cm}C{1.2cm}C{2.3cm}C{1.5cm}R{1.4cm}R{1.3cm}R{1.3cm}}
            Model
                & Cond.
                    & R.F.
                        & Blocks
                            & State Dimension
                                & Channels
                                    & \# Params
                                        & FLOP/s 
                                            & MAC/s\\ 
            \hline
            S4-F-S-16 & FiLM & - & 4 & 4 & 16 & 8.9k & 135.2M & 53.8M\\
            S4-TF-S-16 & TFiLM & - & 4 & 4 & 16 & 30.0k & 155.6M & 53.8M\\
            S4-TTF-S-16 & TTFiLM & - & 4 & 4 & 16 & 10.2k & 141.0M & 56.3M\\
            S4-TVF-S-16 & TVFiLM & - & 4 & 4 & 16 & 11.6k & 138.9M & 55.3M\\
            \hline
            \hline
        \end{tabular}
    }
    \centerline{
        \begin{tabular}{L{3cm}C{7.2cm}R{1.4cm}R{1.3cm}R{1.3cm}}
            Model
                & Signal Chain
                    & \# Params
                        & FLOP/s 
                            & MAC/s\\
            \hline
            GB-C-DIST-MLP & PEQ.sc $\rightarrow$ G.sc $\rightarrow$ O.sc $\rightarrow$ MLP $\rightarrow$ G.sc $\rightarrow$ PEQ.sc & 4.5k & 202.8M & 101.4M\\
            GB-C-DIST-RNL & PEQ.sc $\rightarrow$ G.sc $\rightarrow$ O.sc $\rightarrow$ RNL $\rightarrow$ G.sc $\rightarrow$ PEQ.sc & 2.3k & 920.5k & 4.3k\\
            \hline
            GB-C-FUZZ-MLP & PEQ.sc $\rightarrow$ G.sc $\rightarrow$ O.dc $\rightarrow$ MLP $\rightarrow$ G.sc $\rightarrow$ PEQ.sc & 4.2k & 202.8M & 101.4M\\
            GB-C-FUZZ-RNL & PEQ.sc $\rightarrow$ G.sc $\rightarrow$ O.dc $\rightarrow$ RNL $\rightarrow$ G.sc $\rightarrow$ PEQ.sc & 2.0k & 988.9k & 3.6k\\
            \hline
            \hline
        \end{tabular}
    }
\end{table*}
\setlength{\tabcolsep}{6pt}
\renewcommand{\arraystretch}{1.0}
\setcounter{table}{2}
\begin{table}[]
    \small
    \caption{Test loss for parametric models trained on Multidrive Pedal Pro F-Fuzz. Best model for each architecture shown in \textbf{bold}.}
    \label{tab:results}
    \centering
    \begin{tabular}{l>{\columncolor{gray!20}}ccc}
        \hline
        \hline
        Model
        & Tot. & \footnotesize{$L1$} & \footnotesize{MR-STFT}\\ 
        
        \hline
        TCN-F-45-S-16 & 0.7095 & 0.0217 & 0.6878\\
        TCN-TF-45-S-16 & \textbf{0.4886} & 0.0077 & 0.4809\\
        TCN-TTF-45-S-16 & 0.5324 & 0.0102 & 0.5223\\
        TCN-TVF-45-S-16 & 0.5356 & 0.0115 & 0.5241\\
        \hline
        S4-F-S-16 & 0.7687 & 0.0243 & 0.7444\\
        S4-TF-S-16 & 0.4034 & 0.0075 & 0.3959\\
        S4-TTF-S-16 & 0.3816 & 0.0066 & 0.3749\\
        S4-TVF-S-16 & \textbf{0.3354} & 0.0058 & 0.3296\\
        \hline
        GB-C-DIST-MLP & 1.2104 & 0.0611 & 1.1492\\
        GB-C-DIST-RNL & 1.2531 & 0.0672 & 1.1858\\
        GB-C-FUZZ-MLP & \textbf{0.9303} & 0.0345 & 0.8958\\
        GB-C-FUZZ-RNL & 0.9395 & 0.0355 & 0.9040\\
        \hline
        \hline
    \end{tabular}
\end{table}

\section{Audio Effects Modeling}
To showcase our audio effects modeling framework and evaluate the proposed conditioning methods we train parametric black- and gray-box models of the Multidrive Pedal Pro F-Fuzz, a digital emulation of the Dallas Arbiter Fuzz Face.
Table~\ref{tab:models} shows all models configurations. 
We select TCN and S4 models and evaluate all conditioning methods available (i.e., FiLM, TFiLM, TTFiLM, TVFiLM).
The table shows how TTFiLM and TVFiLM enable implementation of time-varying conditioning with a small overhead w.r.t. FiLM.

We propose two gray-box architectures (GB-DIST and GB-FUZZ) that are extensions of the typical Weiner-Hammerstein model \citep{colonel2022reverse} adopted for distortion modeling, which includes a memoryless nonlinearity in between pre-emphasis and de-emphasis linear time-invariant filters.
We test two nonlinearities: Static MLP (MLP) and Static Rational Nonlinearity (RNL).
While GB-DIST models only use Static Conditional controllers, in GB-FUZZ we opt for a Dynamic Conditional controller for the Offset block, to capture the characteristic dynamic bias shift of fuzz effects.

Models are trained for a maximum of 15k steps using a weighted sum of $L1$ and MR-STFT \citep{steinmetz2020auraloss} losses and the results shown in Table~\ref{tab:results}.
For TCN models, TTF and TVF conditioning perform on par with TF; while for S4 models TTF and TVF outperform TF.
For GB models, regardless of the nonlinearity type, GB-FUZZ achieves better results than GB-DIST, proving the Dynamic controller useful.
Also, RNL in shown to be an effective and efficient alternative to the MLP nonlinearity.

\section{Conclusion}
In this work we presented NablAFx, an open-source framework developed to support research in differentiable black-box and gray-box audio effects modeling. 
Its modular design enables easy configuration of experiments with different architectures, datasets, training settings, and loss functions. 
With logging, plotting, and performance metrics, it simplifies experiment analysis and comparison.
We consider gray-box models as a series connection of DDSP blocks, but this could be generalized using a graph representation.
While black-box models are currently single networks, they could be extended to interconnected networks.
Hybrid models could be introduced to combine black- and gray-box processors, allowing DDSP blocks with known functions alongside neural networks for learning functions.
Moreover, community contributions could help expand our framework in various ways, including new architectures, loss functions, metrics, and more.


\section{Acknowledgments}
Funded by UKRI and EPSRC as part of the ``UKRI CDT in Artificial Intelligence and Music'', under grant EP/S022694/1.

\bibliographystyle{unsrtnat}
\bibliography{references} 

\end{document}